\documentclass[10pt, oneside]{article}   	
\usepackage{graphicx}						
\usepackage{amsmath}
\usepackage{amsfonts}
\usepackage{bbm}
\usepackage{authblk}

\title{Using Recursive Partitioning to Find and Estimate Heterogenous Treatment Effects In Randomized Clinical Trials\thanks{Arun Kuchibhotla and Weijie Su provided important insights as the inferential procedures used in this paper were developed.}}

\author[1,2]{Richard Berk}
\author[2]{Matthew Olson} 
\author[2]{Andreas Buja}
\author[1]{Aur\'{e}lie Ouss}
\affil[1]{Department of Criminology, University of Pennsylvania}
\affil[2]{Department of Statistics, University of Pennsylvania}

\begin{document}
\maketitle

\begin{abstract}
Heterogeneous treatment effects can be very important  in the analysis of randomized clinical trials. Heightened risks or enhanced benefits may exist for particular subsets of study subjects.  When the heterogeneous treatment effects are specified as the research is being designed, there are proper and readily available analysis techniques. When the heterogeneous treatment effects are inductively obtained as an experiment's data are analyzed, significant complications are introduced. There can be a need for special loss functions designed to find local average treatment effects and for techniques that properly address post selection statistical inference. In this paper, we  tackle both while undertaking a recursive partitioning analysis of a randomized clinical trial testing whether individuals on probation, who are low risk, can be minimally supervised with no increase in recidivism. 
\end{abstract}

\section{Introduction} 
Heterogenous treatment effects have long been a feature in the design and analysis of randomized clinical trials. The heterogeneity can be formulated as interaction effects between different randomly assigned interventions (Fisher, 1935: 97-101) or as interaction effects between randomly assigned interventions and pre-experimental or fixed attributes of the study units (Fisher, 1935: 207-213). In either case, there are appropriate analysis techniques when the interactions are anticipated in the research design. Examples include blocking (Cox, 1958: Section 2.3), analysis of covariance (Hinkelmann and Kempthorne, 2008, Chapter 8), and estimation using instrumental variables (Angrist, 2004; Swanson et al., 2018).  However, there are significant complications if heterogeneous treatment effects must be discovered inductively when an experiment's data are analyzed. 

In this paper, we address the discovery and estimation of heterogeneous treatment effects in randomized clinical trials (RCTs) when recursive partitioning is the analysis method of choice.  One challenge is defining  loss functions than can effectively find local average treatment effects. Conventional loss functions may not suffice. Another challenge is proper statistical inference after an empirically-guided search for local effects. The problems are in general well known as ``statistical inference after model selection'' (Leeb and P\"{o}tscher, 2005; 2006; 2008). Work is underway on several different solutions (Berk et al., 2013; Taylor and Tibshirani, 2015; Sun et al., 2016; Barber et al., 2018) that might be usefully ported to randomized clinical trials.  

Our approach draws on past work, summarized below, that uses recursive partitioning (Breiman et al., 1984) to search for heterogeneous treatment effects. However, we reformulate the empirical question being asked, the data partitioning criterion, and how statistical inference is conceptualized. We ask directly what subsets of study units, characterized by their covariate values, have the largest positive or negative treatment effects. Our fitting criterion is the size of those effects when a node is partitioned. Statistical inference encompasses the regression tree as a whole, not just the subset(s) of study units identified by the procedure; we suggest a different loss function and different way to undertake statistical inference. We then apply our results to an RCT testing whether low risk offenders can be efficiently and safely released on probation with virtually no supervision. 

\section{The Potential Outcomes Framework Under Random Assignment}

Consistent with much current practice in statistics, we proceed with the ``potential outcomes'' framework. The potential outcomes framework was originally formulated for a fixed population assigned at random to treatment and control conditions  (Splawa-Neyman et al., 1990; Rubin, 1986; Holland, 1986). Uncertainty in the response comes solely from the random assignment. There are recent extensions to ``superpopulations''  we turn to shortly.

For ease of exposition, we define two interventions: a treatment condition $T$ and a control condition $C$. Each study unit $i$ has a hypothetical outcome if exposed to $T$ or $C$,  $\mathbb{Y}_{iT}$ and $\mathbb{Y}_{iC}$ respectively. Both hypothetical outcomes can vary over study units. At the level of a study unit, the treatment effect is $\tau_{i} \overset{\Delta}{=} (\mathbb{Y}_{iT}) - (\mathbb{Y}_{iC}),$ where the response can be numeric or categorical.  With all study units hypothetically exposed to both $T$ and $C$, the average treatment effect (ATE) is $\tau \overset{\Delta}{=} \mathop{{}\mathbb{E}}(\mathbb{Y}_{T}) - \mathop{{}\mathbb{E}}(\mathbb{Y}_{C})$, where the expectation is taken over the relevant population. 

There is a binary indicator for the assigned treatment. $W_{i} = 1$ if the unit is assigned to the treatment condition, and $W_{i} = 0$ if the unit is assigned to the control condition. A study unit's observed outcome $Y_{i}$, depends on the value of $W_{i}$, realized as a consequence of random assignment. We denote the empirical mean of the experimental group as $\bar{Y}_{T}$ and the empirical mean of the control group as $\bar{Y}_{C}$. Then, the estimated ATE is $\hat{\tau}=\bar{Y}_{T} - \bar{Y}_{C}$. The value of $\hat{\tau}$ is ``global'' if computed over all of the data being analyzed and is ``local'' if computed for a subset of those data.\footnote
{
A local ATE is sometimes denoted by LATE. But that acronym is often associated with particular research settings. One example the need to estimate for an RCT the effect of treatment on those whose treatment actually received changed as a result of some intervention, such as variation in an instrumental variable (Angrist, 2004: C57). Imbens (2010) provides a spirited defense of LATE. Because our research setting is very different, we avoid the acronym.
}

More recently, the potential outcomes framework has been extended to ``superpopulations'' (Imbens and Rubin, 2015, Chapter 7). Study units are seen as IID realizations from a superpopulation that are then randomly assigned to treatment or control conditions as before. This is effectively the same as drawing one random sample from a population to use as the experimental group and another random sample from that population to use as the control group.

There are now two sources of uncertainty, random sampling and random assignment (Berk et al., 2014). Although there are important insights from formulating uncertainty in this manner (Athey and Imbens, 2016), we will proceed somewhat differently. The superpopulation is characterized by a joint probability distribution with random variables $Y$, $W$ and $\textbf{X}
$ for the response, treatment assignment, and covariates respectively. $W$ is independent of the hypothetical outcomes. The joint probability distribution is full rank and has the usual four moments. The definitions and concepts from the earlier fixed population perspective apply, but for features of the joint probability distribution, expectations are now taken over the random variables in that distribution. The data seen by a researcher are random IID realizations from this population. As explained shortly, a form of valid statistical inference follows that can be instructive for finding and estimating heterogenous treatment effects in RCTs. 

\section{Heterogeneous Treatment Effects Using Recursive Partitioning}

Under random assignment, proper estimates $\hat{\tau}$ are most simply obtained from the difference between the mean response from those study units for which the realized $w_{i} = 1$ and those study units for which the realized $w_{i }=  0$. Far more challenging is obtaining valid estimates for $\hat{\tau}$ conditioning on realized values of $X$, where $X$ defines a neighborhood in covariate space where the average treatment effect is systematically atypical.  

\subsection{Recursive Partitioning In General}

Recursive partitioning has been popular approach for finding and estimating heterogeneous treatment effects (Su et al., 2009; Foster et al., 2011; Lipkovich et al., 2011; Loh et al., 2015; Lee et al., 2016;  Athey and Imbens, 2016; Tomas and Bornkamp, 2016). Its popularity is consistent with several well-known assets that apply to recursive partitioning in general. We summarize these assets to set the stage for what follows. In the pages ahead, we use the terms ``partition" and ``node'' interchangeably.  

\begin{enumerate}
\item
The goal of recursive partitioning, whether used with RCTs or not, is to explore regression functionals for the conditional distribution of $Y|X$. This intent is consistent with the aim of finding heterogeneous treatment effects that depend on covariates.
\item
There is no need for a model formulated in advance, and there is no need to treat the results as a model. Recursive partitioning is a \textit{procedure} that adaptively defines covariate neighborhoods in response to some fitting criterion. Although one may choose to impose a modeling perspective, one must then address all of the usual model misspecification concerns. Why step functions? Why a stagewise algorithm? Why the included covariates and not others? 
\item
When covariate neighborhoods of interest are found, their defining x-values are apparent. This allows for subject-matter expertise to help determine how credible the results really are. In addition, one is able to convey what kinds of study units reside in each neighborhood. This is vital if the results are to be used subsequently in real-world situations. One is able to accurately specify the covariate values associated with study units in the target neighborhood (e.g., males under 25 years of age with a high school education). This facility cannot be obtained from  ensembles of trees like random forests or gradient boosting because for each tree there typically will be different covariate neighborhoods.
\item
The idea of recursive partitioning is very general and has been implemented in many different and creative ways. The procedure can be hand-tailored with relative ease as the needs and insights of researchers require.
For any internal partition, there can be a subsetting into a left $(L)$ child partition and a right $(R)$ child partition. There are $n_{L}$ study units in the left partition and $n_{R}$ in the right partition. Within each, a summary statistic is computed, such as a mean. A variety of summary measures can be used.  There are associated disparities between summary statistic and the observed values of the response that can aggregated. The error sum of squares (MSE) is one example. There are many others. For binary response variables, the Gini is a popular option. From these, an overall measure of fit is obtained such as mean squared error. Such fit measures are often called the ``loss.'' The quality of a particular partitioning is measured by the reduction in loss compared to the loss for the parent partition. For split $S$, parent node $A$, and child nodes $B_{L}$ and $B_{R}$, the reduction in loss can be expressed very generally as
\begin{equation}
\Delta {\rm Loss} (S, A) = {\rm Loss}_{A} - p({\rm Loss}_{B_{L}}) - ((1-p)({\rm  Loss}_{ B_{R}})),
\label{eq:delta}
\end{equation}
where $p$ is the proportion of cases falling in the left child partition, and  $(1-p)$ is the proportion of cases falling in the right child partition.
\item
Each potential split for each predictor can be represented as an indicator variable coded ``1'' the the right of the split value and ``0'' at the split value or below it.  One can then represent the full tree as a single linear regression with the indicator variables for regressors. In this form, it can be easier to compare the results to other analyses using linear regression. 
\item
The ``greedy'' algorithms conventionally applied work well in practice, although there are rarely any performance guarantees. Reliance on a stagewise approach limits the population of recursive partitionings that can be explored -- one does not get to evaluate all possible tree structures. In addition, the use of step functions, which are necessarily discontinuous, complicates the derivation formal inferential properties. 
\item
Recursive partitioning can be used for regression functionals characterizing the conditional distribution of $Y|X$ in different ways (e.g., parametric versus nonparametric). These functionals are taken to be features of a joint probability distribution and estimated with IID realized data from that joint probability distribution. Buja and his colleagues (2018a; 2018b) provide a wide ranging, detailed discussion of regression functionals in a variety of settings.
\end{enumerate}

Unfortunately, the ways in which recursive partitioning determines its regression functionals come with several widely-recognized difficulties (Berk, 2016: Chapter 3). Perhaps most important, the neighborhoods are adaptive. As the recursive partitioning unfolds, the procedure searches at each stage over all possible splits for all available predictors to find the ``best'' pair of data subsets. One has a form of automated ``data snooping'' that risks serious overfitting and as a method of variable selection can, as noted above, invalidate statistical inference (Leeb and P\"{o}tscher, 2005; 2006; 2008). 

In addition, the data partitions chosen can be very unstable. If there are two or more candidate partitionings that perform about equally well, the chosen partitioning may perform only slightly better than its competitors. With new realizations of the data, another partitioning may well be chosen instead. Because the algorithm proceeds in a stagewise manner, instability in the first few partitions cascades through later partitions. 

Finally, the data do double duty.  They are used to arrive at the recursive partitions \textit{and} to estimate various features of the response distribution within each. This carries the damaging estimation consequences of the automated data snooping into all of the estimated regression functionals. Ideally, there are other data, not used to determine the partitioning, that can be used for the estimation step. A random, holdout sample is one instance. More recent machine learning procedures, such as random forests, capitalize on this idea (Breiman, 2001; Athey et al., 2017). Later, we provide another approach.

\subsection{Finding Heterogenous Treatment Effects}

Recursive partitioning is easily altered to search for local ATEs. The aim is to find one or more locations in the covariate space in which the disparity between the global ATE and the local ATE is large. A reasonable summary measure is the within-partition ATE with the global ATE subtracted: ATE$^{\ast}_{local} \overset{\Delta}{=}  (\bar{Y}_{T_{local}} - \bar{Y}_{C_{local}}) - (\bar{Y}_{T_{global} } - \bar{Y}_{C_{global}})$, where the asterisk indicates a form of centering. Because the global ATE is a constant, the centering has no impact on the partitioning process. It does, however, affect how one decides whether a local ATE is important in subject-matter or policy terms. It also affects the test statistic we use later.\footnote
{
In some situations, some might favor using the null hypothesis value for the global ATE. For example, if one has already failed to reject the the null hypothesis that ATE$^{\ast}_{global} = 0.0$, one might argue for using 0.0 as the global ATE. The risk is that one is proceeding as if a statistical test already has shown that ATE$^{\ast}_{global}$ actually is zero. Using the estimated global ATE to center the local ATE's is the option we prefer, although that introduces another source of uncertainty. Fortunately, because the global ATE is estimated from the full set of observations, the amount of uncertainty introduced may be very small, especially compared to the uncertainty in the local ATE estimates. Still, one could address that small uncertainty with bootstrap extensions of the procedures we later provide.  
}

${\rm \widehat{ATE}}^{\ast}_{local}$ is computed for each left and right prospective partition from which a loss function is constructed. For example, two separate within-partition linear regressions can be run taking the form of $Y = \beta_{0} + \beta_{1}W + \varepsilon$.  $W$ is coded as 1 or 0. ${\rm \widehat{ATE}}^{\ast}_{local}  = (\hat{\beta}_{1} - \hat{\beta}_{0}) - {\rm \widehat{ATE}}_{global}$, which can be positive or negative. Each regression error sum of squares (ESS) is computed in the usual manner. One can then calculate the combined MSE for the two partitions as before. A small MSE is desirable. Alternatively, one might compute a test statistic for the null hypothesis that $\beta_{1} = 0.0$. A large value for the test statistic is desirable. To date, there seems to have been several interesting variations on these approaches (Su et al., 2009; Lipkovich et al., 2011; Loh et al., 2014; Athey and Imbens (2016). We will have more to say the interesting work by Athey and Imbens below. In short, conventional recursive partitioning with numeric response variables aims  to reduce average heterogeneity at each potential split. The two partitions chosen have the smallest weighted average of their within-group heterogeneities.\footnote
{
Although beyond the scope of this paper, there are also model based methods for finding and estimating heterogeneous treatment effects. Some come from economics (Angrist, 2004). But statisticians can like models too. Thomas and Bornkamp (2016) specify a suite of models in which beyond an indicator for the intervention, one includes the main effects of covariates and interaction effects between the covariates and the intervention. Both the main effects and interaction effects can be subject to different transformations from model to model. Nonparametric regression (e.g., cubic regression splines) is one example. Statistical tests are then used to determine which interaction effects are retained. To deal with the multiplicity of models, the authors suggest model averaging or beginning with one all encompassing model, undertaking dimension reduction using the lasso.  
}

We offer a rather different and novel formulation based on work by Buja and Lee (2001). Rather than searching indirectly for heterogenous treatment effects using data partitions that most improve a measure of fit, we search directly for partitions that have atypically large positive or negative average treatment effects. In many policy settings, one cares especially about the subset of study subjects for whom the intervention works unusually well or unusually poorly. 

For each possible splitting variable and each possible splitting value for each variable, the centered local average treatment effect is calculated separately for the two potential data partitions. For the left (L) and right (R) partitions, one determines $max({\rm \widehat{ATE}}^{\ast}_{local(L)}, {\rm \widehat{ATE}}^{\ast}_{local(R)})$. This is repeated over all possible predictors and splits.  The largest $max({\rm \widehat{ATE}}^{\ast}_{local(L)},$  ${\rm \widehat{ATE}}^{\ast}_{local(R)})$ over all possible predictors and splits decides the ``best'' partitioning.\footnote
{
Lipkovich and colleagues consider a related idea using test statistics instead  (2011).
}
Alternatively, one can determine $min({\rm \widehat{ATE}}^{\ast}_{local(L)}, {\rm \widehat{ATE}}^{\ast}_{local(R)})$ This is repeated over all possible predictors and splits.  The smallest $min({\rm \widehat{ATE}}^{\ast}_{local(L)},$ ${\rm \widehat{ATE}}^{\ast}_{local(R)})$ over all possible predictors and splits decides the ``best'' partitioning. One should chose the max-rule or the min-rule (or both) depending on subject matter or policy concerns. For example, is one looking for the study units whose average treatment effect is most beneficial or most harmful (or both)? 

We employ as a tuning parameter the minimum  number of observation for a covariate neighborhood. In effect, one is tuning for a good bias-variance tradeoff.\footnote
{
One can also tune for a good bias-variance tradeoff by stopping the partitioning after a very small number of splits or by pruning. 
}
The well-known instability of recursive partitioning (Berk, 2016: Chapter 3) underscores the need to think long and hard about reducing the variance.  Tuning for partition size can also make the results more responsive to policy concerns. One might prefer a weaker heterogeneous treatment effects if the number of study units in the covariate neighborhood is larger. More study units can be served in the future even though the expected benefits for each may well be smaller on the average.

We favor the $max$ or $min$ splitting criteria because each directly addresses what we are seeking. We want to find neighborhoods in covariate space where the study units perform toward either tail of the local ATE distribution. We are not especially interested fit quality or choosing splits to reduce some measure of fitting error. These can be useful goals, but are not a high priority for our problem.

\subsection{Statistical Inference for Heterogenous Treatment Effects}

As noted above, recursive partitioning is adaptive. Partitions are defined through searches over predictors and splits of those predictors. The process unfolds in a stagewise fashion so that earlier partitions are not reconsidered as later partitions are contemplated. Consequently, the impact of inductive variable selection compounds. Statistical inference for any summary statistic computed for a response in a given covariate neighborhood risks highly undesirable inferential properties. 

\subsubsection{Honest Statistical Inference}

One promising and widely appreciated improvement exploits data, realized in the same fashion from the same joint probability distribution, subject to the same interventions randomly assigned, but not used to determine the recursive partitions (Loh, 2011). Such ``test data'' can be used to populate the terminal nodes -- one can say that the test data are ``dropped'' down the regression tree constructed from the partitions until each observation lands in its appropriate tree leaf. Summary measures are computed, and statistical inference is undertaken in any terminal node using solely the test data in that terminal node. Then, valid statistical inference can proceed as usual. In practice, researchers often favor dividing a dataset into two random, disjoint sets: training data to determine the partitions and test data to compute statistical summaries.

But data splitting forces tradeoffs. Statistical precision is necessarily lost because the number of observations in both the training data and the test data is substantially reduced compared to the number of observations in the entire dataset. In addition, data splitting can increase bias because recursive partitioning is sample-size dependent. With smaller samples, one grows smaller trees because the algorithm exhausts all available observations sooner. A smaller number of terminal nodes can increase bias in the fitted values. Finally, although working from a single split is easy to implement, it can make the results vulnerable to the one-trial subsetting of the data. Meinshausen and his colleagues (2009) call this a ``p-value lottery'' because it can be difficult to reproduce results. They favor a more demanding form of statistical inference from multiple splits whose results can be aggregated, although their solution is limited to variable selection for conventional linear regression that has predefined predictors.\footnote
{
By ``pre-defined,'' one means that all of the potential regressors are defined before the data analysis begins. They are not constructed as part of the data analysis, which is precisely what recursive partitioning does (i.e., indicator variables defining splits). For regression, their method combines a powerful variable selection procedure with proper controls for both the family-wise error rate and the false discovery rate. 
}

It can also be desirable to adjust for the potential bias from the number of possible splits that a covariate can have (Hothorn et al., 2006). Covariates with a larger number of possible splits have more opportunities to be selected. Such covariates can give the false impression that they are important splitting variables than they actually are not.\footnote
{
Although this can be important if one is going to use the tree-structure for subject-matter explanations, less clear is whether it matters for estimates of local ATEs, especially given our preferred loss functions. For subsequent use, the data will be random IID realizations from the same joint probability distribution that includes the same variables measured in the same way. 
} 

Finally, because recursive partitions are determined adaptively, the summary statistic used to determine the partitioning can be optimistically biased by the data snooping. Applying K-fold cross-validation to the fit measure can help, although there remain a number of unanswered questions about the properties of cross-validation and how it is best used (Hastie et al., 2009: Section 7.10). There are, for instance, important bias-variance tradeoffs that result from the choice of K. 

Athey and Imbens (2016) draw on such ideas and call the resulting statistical inference ``honest.'' A key to their approach is that a random split of the data is used for fitting, and the data complement is used for estimation and inference. This has many desirable features of traditional statistical inference. For such inference, all selections and transformations of random variables for $Y|X$ are decided \textit{before} the data analysis begins. The variable selections and transformations can be determined by prior research, scientific theory or even researcher whimsey. The point is that the selections and transformations are not influenced by an analysis of the same data used for statistical inference. Routine statistical inference properly can follow. 

But honest statistical inference does not incorporate uncertainty from anything that happened before one or more terminal nodes are populated with test data. It conditions on the training data, the random division of the dataset into training and test data, and the resulting recursive partitions. The only data analyses for which uncertainty is allowed are the estimates of any local ATEs and subsequent statistical tests and confidence intervals. All that transpired earlier is treated as fixed, including which subgroups appear to have important local treatment effects in the training data. 

Yet, the prior steps and the estimation with test data all depend on the \textit{same} experiment. How that experiment is analyzed, including the recursive partitioning, affects how the test data are generated. In particular, the candidate subgroups for heterogeneous treatment effects determine which test data observations used (i.e., only for the subgroups chosen with the training data). Stepping back farther, had the experiment been done with a different set of IID observations from the same population, each step in the analysis almost certainly would have had different results, including the estimation of local average treatment effects. In short, important sources of uncertainty are conditioned away. 

One important consequence is that honest statistical inference risks playing into current concerns about the reproducibility of science (McNutt, 2014; Baker, 2016; Johnson, 2016). Honest statistical inference is agnostic with respect to what might happen if a new experiment were conducted: a new set of study subjects, a new random assignment process, a new estimate of the overall ATE, a new random split of the data, a new recursive partitioning, and new local ATE estimates for the subgroups found. Reproducibility may not matter to a policy maker needing to act on the best information currently available, but is an essential consideration for determining in general  ``what works.''

One might respond that for the particular application at hand, all that matters is the uncertainty in the test data. The earlier sources of uncertainty are irrelevant. Such a rationale would need to be situation specific and carefully articulated. For example, policy makers who plan to use the results might be satisfied treating the selected subgroups as the right ones because they believe that the uncertainty ignored would not materially change the results. Going forward, they hold that the intervention will be better delivered. If their rationale can be sufficiently justified, the results become ``good enough,'' at least until another study can be mounted. For example, it would help if for an exercise using different random samples of the training data with replacement, recursive partitioning always produced the same terminal nodes. In that case, the instability from recursive partitioning over samples would not appear to be a problem. 

Honest statistical inference perhaps is best justified more generally if it is the best one can do. Proper statistical inference with inductive data analysis procedures is challenging and for many applications, the problems are unsolved. Honest statistical inference can offer partial solutions and is certainly better than using the same data for fitting and for estimation.  But, in the search for heterogeneous treatment effects using recursive partitioning, there is an alternative approach that captures a far larger portion of the uncertainty. We turn to that now. 

\subsubsection{Righteous Statistical Inference}

We offer an alternative that one might call ``righteous,'' building on recent work addressing post selection inference (Berk et al., 2013). One can achieve proper control over the family-wise error rate for the very large set of possible partitions that might be selected. There is, therefore, effective protection against false discoveries, which otherwise would be quite likely. 

The estimand is the local ATE for specified neighborhoods in the covariate space represented in the joint probability distribution responsible for the IID data. An example might be men 25 years of age, with 3 prior arrests for drug possession. We are interested in the local ATE for such study subjects.\footnote
{
The connection to model selection becomes more apparent when one recalls that each split can can be represented by an indicator variable. For each possible split, one seeks to find the best indicator variable, discarding all others. 
} 

Both the neighborhoods and the local ATEs are determined by an inductive search over the data. In this paper, we focus on recursive partitioning. Righteous statistical inference requires that one enumerate empirically \textit{all possible data partitions} that could have been produced over new realizations of the data, conditional on tuning parameters values, such as the minimum number of observations in a partition. The local ATE in each such partition of the data is then an unbiased estimate of the population local ATE for the corresponding covariate space.  Consider, for example, a single categorical regressor with three levels (i.e., a, b, c). There are 6 possible splits (i.e., a, b, c, ab, ac, bc), each with its own ATE in the population and estimated from the data. All such splits for all regressors must be considered. 

We suggest the partitions be constructed from stumps with some extensions, described later, to consider trees of greater depth. Because the number of data partitions grows extremely fast with tree depth (i.e. exponentially), even depth-2 partitionings with several predictors can become a computational challenge.\footnote
{
For example, if there are 400 possible partitions for each pass through the data, a depth-1 partitioning has 400 possible partitions. A depth-2 partitioning as 160,000 possible partitions. A depth-3 partitioning has over 64 million possible partitions. And 400 depth-1 partitions is not at all extreme. Suppose a single categorical variable, such as the kind of crime for which a person was arrested, has 8 possible categories. There are 254 possible partitions from that covariate alone.
}
With stumps, a full enumeration of all possible partition usually is quite feasible.

The standard null hypothesis is that the difference between the global ATE and each local ATE is zero: ${\rm ATE}^{\ast}_{local} = 0.0$ in each  partition. Under the null hypothesis, therefore, each partition is equally likely to be generated in the search for heterogeneous treatment effect. There is no need to weight the partitions for their probability of occurrence. Our loss function relies on local ATEs as its sole argument, and we use conventional t-scores as our test statistic.

Random assignment makes the intervention independent of all covariates. A simple permutation procedure naturally follows. The idea of permutation inference can be traced back to Fisher (1935) with formal theory provided by Kempthorme (1955) among others. One can repeatedly permute the label for the assigned intervention to obtain an empirical approximation of a test statistic's distribution under the null hypothesis (Edgington and Onghena, 2007: Chapter 3). For any estimated local ATE from the earlier recursive partitioning, one then can determine the probability under the null hypothesis that one could obtain in new realizations of the data an estimated local ATE for any partition as large or larger than the one obtained from the recursive partitioning. This procedure provides reasonable insurance against the family-wise error rate, while properly incorporating uncertainty from IID data, the random split of the data, and the partitioning process.

The null distribution is agnostic about how the recursive partitioning is done. For example, the same rationale applies for any of several different recursive partitioning loss functions and algorithms including Quinlan's ID3 (1986) and Kass's CHAID (1980).  More generally, our permutation procedure provides valid inference for \textit{any} selection procedure one might consider (Berk et al., 2013).  Recursive partitioning, whether using stumps are not, is a special case. 

Let $\mathcal{R}$ denote the set of possible splits defined by the covariates $X$, $t_{R}$ the associated t-score for a given split $r \in \mathcal{R}$.  Then for any potential selected split $r^*$, 
\begin{equation*}
P(t_{r^{*}} \geq c) \leq P(max_{r \in \mathcal{R}} t_r \geq c) \leq \alpha,
\end{equation*}
where the probability is computed under the null (permutation) distribution of no association.  In other words, we compute the null distribution for a statistic that strictly dominates any specific selected statistic, providing valid, conservative inference. However, the results can be sobering. Righteous statistical inference makes painfully plain how much uncertainty is really in play.

In practice, one can proceed as follows.
\begin{enumerate}
\item
Examine the marginal distribution of each covariate. Consider binning the numeric variables into, say, quartiles, and collapsing some rare classes for categorical variables. Rare values for either will not make the cut for a reasonable minimum number of observations in a partition. One can achieve computational and interpretive simplification. 
\item
Apply recursive partitioning to the full dataset using the maximum ATE and/or minimum ATE loss. We recommend regression stumps.
\item
Select one or more partitions of interest and compute t-values for null hypothesis that the difference between the global ATE and the local ATE = 0.0.
\item
Compute an empirical approximation of the null distribution for the maximum and/or minimum t-value. For example, if the treatment indicator is permuted 1000 times, there will be 1000 maximum or minimum t-values under the null hypothesis that the difference between the global ATE and all local ATEs, defined by all possible splits over all possible covariates, is equal to zero.\footnote
{
Across all possible splits of the data for a specified, minimum number of terminal node observations, there will be one maximum t-value or one minimum t-value each time the treatment indicator is permuted. 
}
\item
Repeat steps 2 - 4 for each tuning parameter value used and accumulate all of the maximum and/or minimum t-values.
\item
Estimate from the empirical null distribution the probability of obtaining a local t-value as large or larger or as small or smaller than the ones obtained from the recursive partitioning. There will be t-values for each value of whatever tuning parameters are manipulated. These are simply combined.
\item
If those probabilities are sufficiently small, consider rejecting the null hypothesis. 
 \end{enumerate}
 
In practice, the stump approach can find splits at depths greater than 1. One simply specifies in advance interaction effects of interest as new covariates to be included among the original covariates. For example, one might include gender and education as main effects and their product (if coded as numeric variables) to capture their interaction effect. Perhaps the intervention is especially effective for women with only a high school education.\footnote
{
Products of numeric variables can be challenging to interpret because there will often be several different values of constituent covariates that have the same product. Binning can help. Then one can more easily define indicator variables to be used as covariates that define particular splits (e.g.,  women under 30 years of age with 12 years of education). It is important to keep in mind that there is no model, and the enterprise is not explanation. We are searching for identifiable subsets of study units with unusually large or small ATEs. 
} 

Another extension is to fit a stump several times, each time dropping the previously selected best splitting variable. One properly can find a second best subset, a third best subset and other bests in this manner. There may be several groups with important heterogeneous treatment effects. 

Finally, because recursive partitioning can be so unstable, it is useful in practice to require that any partition have a substantial number of observations. The minimum number of observations allowed in a partition becomes a tuning parameter that applies to righteous inference and to recursive partitioning used to find any heterogeneous treatment effects. Then, thanks to random assignment, there will be roughly the same number of cases exposed to the treatment condition and the control condition within each partition.\footnote
{
This assumes that each of arm of the experiment is assigned with the same probability. If not, within each partition. the unequal assignment probabilities will be approximately reproduced.
}
Except in highly unlikely cases, one can compute the local ATE as usual. 

In summary, righteous statistical inference applied to the search of heterogeneous treatment effects from an RCT provides valid statistical inference when one has proceeded in an inductive manner, and one is concerned about reproducibility. Righteous statistical inference directly confronts what might happen if an RCT were mounted again from scratch. But taking proper account of the uncertainty inherent in replicating an RCT will generally (and properly) increase standard errors and other measures of chance variability. 

\subsubsection{Implications for Practice}

Because honest statistical inference treats the selected subgroups as fixed, one proceeds as if the subgroups selected by recursive partitioning are at least instructive and ideally, the right ones. For these subgroups, one can test whether their local ATEs are zero in the joint probability distribution. For example, a subgroup of interest might be males, less than 21 year of age, with two prior probation sentences. For that subgroup, honest inference is easily implemented using the test data. 

Righteous statistical inference recognizes that the subgroups chosen could have been different. In another experiment with IID observations from the same population, the recursive partitioning might have chosen, say, the women, with a college education, who are unemployed. Therefore, uncertainty in the subgroups selected must be incorporated into all statistical inference. Test data by themselves are insufficient. Post-selection inference using the full dataset is required. 

In practice, one can apply both honest and righteous statistical inference. For those who are prepared to bet on the subgroups chosen, honest statistical inference is valid. For those who want to allow for the subgroups chosen to have been different, righteous statistical inference is valid.

\section{An application}

Supervising offenders on probation is expensive. Ideally, scarce resources should be allocated to those offenders who most need them. In 2007, the Philadelphia Adult Department of Probation and Parole (APPD) adopted a machine learning risk assessment tool to distinguish between offenders forecasted to be arrested for a violent crime, offenders forecasted to be arrest for crime that was not violence, and offenders, labeled ``low risk,'' who where forecasted to not be arrested for any crime. A little more than half of the overall case load was forecasted to be low risk. Supervisory resources were to be reallocated away from the low risk offenders to those who posed a greater threat to public safety. 

But would the less intense level of supervision increase the number of low risk offenders who were arrested for new crimes? To answer this question, a randomized clinical trial was undertaken. Starting with October 1st, 2007, 1,559 low risk probationers, who had been under supervision for less than three months and who had at least one month still to serve, were randomly assigned to one of two interventions (details in  Ahlman and Kurtz, 2009 :5). The experimental group was assigned to probation officers with cases load of approximately 400 low risk offenders. By design, it was simply impossible to provide as much supervision as previously, and substantial cost saving followed. Office visits were schedule for once every six months with one phone call-in report about midway in between visits. Drug tests was administered only if required by a court order. Arrest warrants were issued if there had been no case contact for more than six months. The controls were assigned to ``standard supervision." Case loads were around 150. Probationers were to report to the main office once a month, but could be required to report as often as once a week. Field visits were unusual, but in some cases, offenders were visited in their residences. If an offender failed to report or make contact within 90 days, an arrest warrant would generally be issued. 

Each study subject was followed through local county criminal justice records for a full 12 months. All arrests were recorded. Several forms of analysis led to the same conclusion: the number of arrests was no greater for offenders who received virtually no supervision. Public safety did not suffer, and there could be a more efficient use of tax dollars. In response, a substantial reorganization of APPD was undertaken that remains in place to date. (See Berk et al., 2010 for a complete discussion of the experiment.)

About 18\% of all offenders in the study were re-arrested within 12 months. But the global ATE over all  study subjects was effectively 0.0 ${\rm(\widehat{ATE}}_{global} = .014$). We wondered if there were subsets of probationers for whom their ATE$_{local}$ was meaningfully positive or negative. A negative local ATE would mean that the experimental group has fewer arrests than the control group. A positive local ATE would mean that the experiment group had more arrests than the control group. We obtained the study data to examine both possibilities. 

Ten covariates were chosen as candidates for splitting.
\begin{enumerate}
\item
Date of Birth
\item
Gender
\item
Race
\item
The age of the earlier charge as an adult
\item
The date of the most recent prior charge
\item
The age as the start of the probation sentence
\item
The number of prior charges for drug possession
\item
The number of prior charges for any crime
\item
The number of prior probation sentences
\item 
From subject-matter knowledge, a product variable for an interaction effect between gender and the number of prior probation sentences, as one might find from a recursive partitioning with depth equal to 2. (The variable was designed to capture whether the local treatment effect for offenders will many prior probation sentences differed for men and women.)
\end{enumerate}

We tried to be judicious in the number of predictors chosen, because with a larger number of predictors, there are more possible splits creating more uncertainty. We also binned each numeric covariate into 10 equally wide intervals. Based on subject-matter knowledge, the loss of information was probably too small to matter, and the number of potential splits was reduced substantially. 

A stump regression tree was fit to the data using the max-ATE or min-ATE loss functions described earlier. From experience with other datasets, we anticipated that for reasonable stability at least 100 observations would be needed in any data partition. We tuned for minimum partition sizes of 100, 150, 200 and found the largest results using 100 observations. The same splitting variable was selected for each, but the split value changed somewhat. With larger minimum bucket sizes, the mix of offenders became more diverse and the local ATE was diluted.  

The 116 Offenders with more than 6 prior probation sentences were 16.5 percentage points more likely to be rearrested under the experimental condition than the control condition. With a global ATE of 0.014, the local ATE is dramatic, and the re-arrest base rate was nearly doubled. They were much worse risks under reduced supervision than the average low risk offender.\footnote
{
For minimum partition sizes of 150 and 200, the split values were substantially lower. Less troublesome offenders were being thrown into the mix.
}
For them, the intervention could be seen as ill advised. Past failures on probation are a strong indicator of future failures on probation especially when supervision is diluted.\footnote
{
The interaction effect with gender, representing a tree depth of 2, was not selected.
} 

Second best and third best partition were found, as described above, by dropping earlier best partitions found and re-running the analysis. Individuals who began their criminal activities at a young age were more likely to fail under reduced supervision compared to standard supervision. But the effect was modest. The gain in re-arrest was only 6 percentage points, perhaps because they had not failed many times on probation before. Still, it is well known that offenders who start young are more likely to fail on probation. 

We undertook a similar analysis for the smallest (i.e., most negative) average treatment effects. We hoped to find one or more subsets of offenders who thrived under reduced supervision compared to standard supervision. We found no partitions for which the ATE was negative and large (i.e., the treatment group was substantially less likely to be rearrested).  

\begin{figure}[htbp] 
   \centering
   \includegraphics[width=4in]{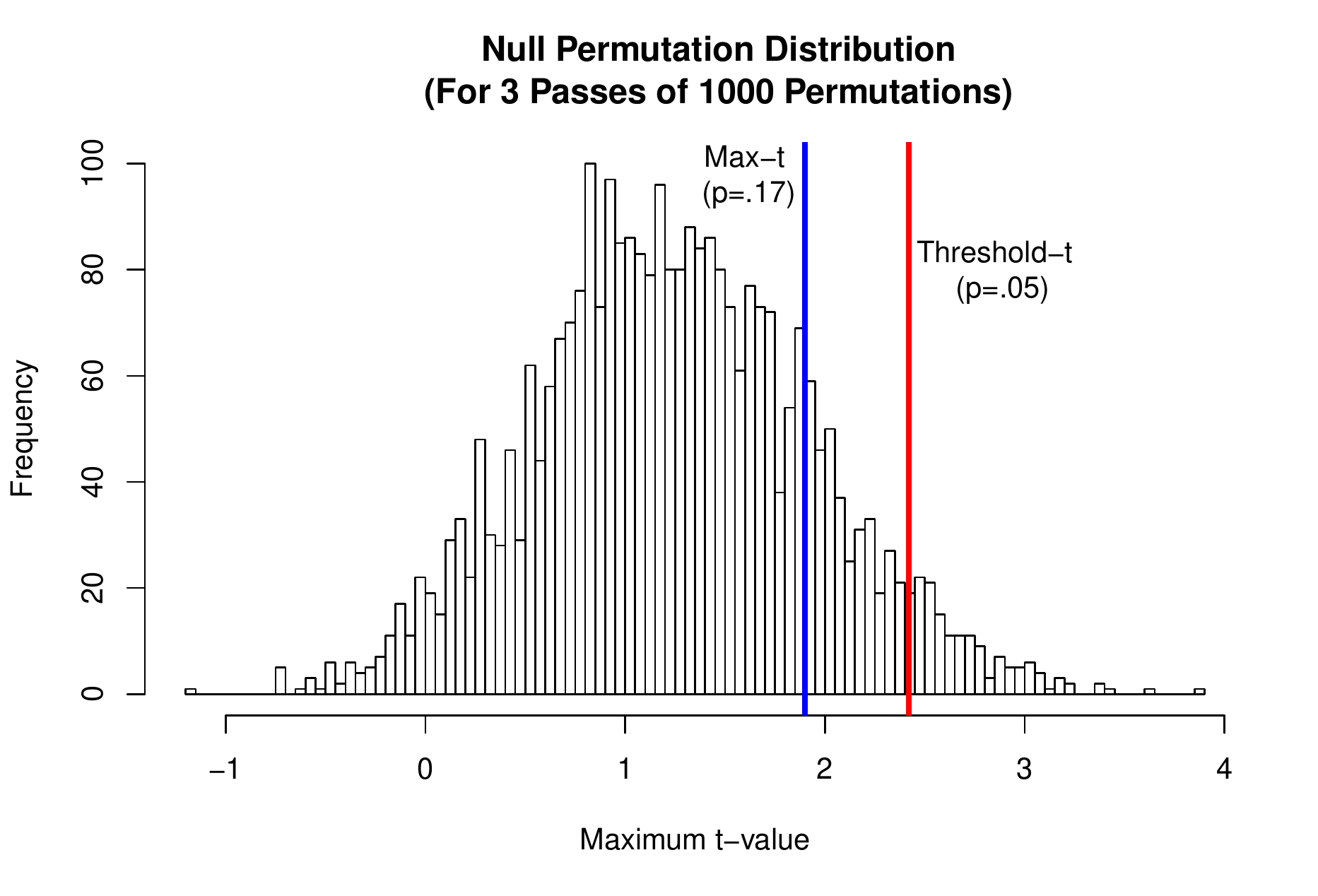} 
   \caption{Exhaustive Null Distribution of Maximum t-Values (ATE$_{local} = 0.0$) For All Possible ATE Partitions With at Least 100 observations (3 passes over the data for 3 different minimum subgroup sizes -- maximum t-value found in blue, threshold t-value in red)}
   \label{fig:null}
\end{figure}

Had a conventional statistical test been undertaken for the recursive partitioning output for the largest local ATE (i.e., ATE$_{local} = .165$), the null hypothesis for a one tailed test would have been rejected with a p-value of .03. But this would be neither honest nor righteous. Statistical inference would have been undertaken employing the same data used for the recursive partitioning with none of the requisite adjustments righteous statistical inference provides.  
 
Righteous statistical inference was applied for three passes through the data for the three different minimum partition sizes used in tuning (i.e., 100, 150, 200). Figure~\ref{fig:null} shows the null distribution including all three passes through the data.  The t-values along the horizontal axis cluster between about  0.5  and 2.0. Righteous statistical inference yields a p-value of .17 for the maximum ATE of .165. A maximum t-value of 2.85 would have been required to pass the .05 threshold.

The p-value of .17 was dramatically affected by the tuning values chosen. The largest local ATE was found with a minimum partition size of 100 observations. Because our intent was to find the largest local ATEs, the results we have stressed correspond to that partition. For minimum partition sizes of 150 and 200, the local ATEs and maximum t-values were somewhat smaller. Despite the larger numbers of observations, smaller t-values pulled in the tails of the permutation distribution leading to greater power. In advance, there was no way to know whether searching over three different minimum partition sizes would increase or decrease our test's  power. 

We do not report righteous statistical inference for the partitions with positive ATEs. The ATEs are too small to matter in practical terms and would not lead to a rejection of the null hypothesis. For example, offenders who were first arrested after the age of 45 were about 5 percentage points less likely to be re-arrested while on probation/parole. Whether or not that is an important practical difference, righteous statistical inference would not have led to a rejection of the null hypothesis.

There are important lessons for test data. If we had test data with a little over 100 individuals who had more than 6 prior probation sentences, we could have populated the best partition with data that were uncontaminated by the recursive partitioning search algorithm.\footnote
{
To actually have that many such individuals would have required test data with about 1500 observations. 
}
It is very likely the same overall conclusions would have been reached. With honest statistical inference, one would have rejected the null hypothesis. With righteous statistical inference one would not. To reject the null hypothesis using righteous statistical inference would have required a far larger number of test data individuals with more than 6 prior probation sentences, assuming the local ATE computed would not change much. 

\section{Conclusions}

Recursive partitioning can be a useful procedure in the search heterogeneous treatments effects from randomized clinical trials. But, there are inherent complications. One needs to consider whether the usual loss functions are sufficiently responsive to the questions being asked and if not, to employ a more responsive alternative. In addition, the automated search over split values and predictors is a form of data snooping that introduces substantial uncertainty and invalidates conventional statistical inference. In response, we proposed Max-ATE or Min-ATE loss combined with permutation inference. The former provides a direct way to seek large heterogeneous treatment effects. The latter offers valid statistical inference. We also emphasize that the choice of loss function and inferential approach depend on the research setting and the questions being asked. There can be legitimate reasons for taking different paths. However, whatever the path taken, clear reasoning must be provided. 

\section*{References}
\begin{description}
\item
Ahlman, L.C. and E.M. Kurtz (2009) ``The APPD Randomized Controlled Trial in Low Risk Supervision: The Effects on Low Risk Supervision on Rearrest.'' Philadelphia: Adult Probation and Parole Department.
\item
Angrist, J. (2004) ``Treatment Effect Heterogeneity In Theory And Practice.'' The Royal Economic Society Sargan Lecture, \textit{The Economic Journal} 114(494) C52--C83.
\item
Athey, S., and Imbens, G. (2106) ``Recursive Partitioning For Heterogeneous Causal Effects.'' \textit{Proceedings of The National Academy of Sciences} 112(27): 7353--7360.
\item
Athey, S., Tibshirani, J. and Wager, S. (2017) ``Generalized Random Forests'' arXiv: 1610.0127v3 [Stat.ME].
\item
Baker, M., (2016) ``1,500 Scientists Lift The Lid On Reproducibility.'' \textit{Nature} 533: 452--454.
\item
Barber, R.F., Cand\`{e}s, E.J., and Samworth, R.J. (2018) ``Robust Inference With Knockoffs.''
 arXiv:1801.03896v3 [stat.ME].
 \item
Berk, R.A. (2016) \textit{Statistical Learning from a Regression Perspective,} Second Edition. New York: Springer.
\item
Berk, R.A., Barnes, G., Alhman, L., and Kurtz, E. (2010) ``When Second Best Is Good Enough: A Comparison Between A True Experiment and a Regression Discontinuity Quasi-Experiment.'' \textit{Journal of  Experimental Criminology} 6(2): 191--208.
 \item
Berk, R.A., Brown, L., Buja, A., George, E., Zhang, K., and Zhao, L. (2013).``Valid Post-Selection Inference.'' \textit{The Annals of Statistics} 41(2): 802--837.
\item
Berk, R.A., Brown, L., Buja, A., George, E., Zhang, K., and Zhao, L. (2014) ``Covariance Adjustments For The Analysis of Randomized Field Experiments.'' \textit{Evaluation Review} 34(3-4) 170--196.
\item
Breiman, L. (2001) ``Random Forests.'' {\it Machine Learning} 45: 5--32. 
\item
Breiman, L., Friedman, J.H., Olshen, R.A., and C.J. Stone, (1984) \textit{Classification and Regression Trees}. Monterey, CA: Wadsworth
Press.
\item
Buja, A., and Lee, Y-S. (2001) ``Data Mining Criteria For Tree-Based Regression And Classification.'' Proceedings of KDD, 27--36.
\item
Berk, R.A., Brown, L., Buja, A., Zhang, K., and Zhao, L. (2013) ``Valid Post-Selection Inference.'' \textit{Annals of Statistics} 41(2): 401--1053.
\item
Buja, A., Berk, R.A., Brown, L., George, E., Pitkin, E., Traskin, M., Zhao, L., and Zhang, K. (2018a) ``Models as Approximations, Part I --- A Conspiracy of Nonlinearity and Random Regressors in Linear Regression.'' \textit{Statistical Science}, forthcoming with discussion.
\item
Buja, A., Berk, R.A., Brown, L., George, E., Pitkin, E., Traskin, M., Zhao, L., and Zhang, K. (2018b) ``Models as Approximations, Part II--- A Conspiracy of Nonlinearity and Random Regressors in Linear Regression.'' \textit{Statistical Science}, forthcoming with discussion.
\item
Edgington, E.S., and Onghena. P. (2007) \textit{Randomization Tests} New York: Chapman \& and Hall.
\item
Fisher, R.A. (1935) \textit{The Design of Experiments} New York: Hafner Press
\item
Foster, J.C. Taylor, J.M.G., and Ruberg, S.J. (2011) ``Subgroup Identification From Randomized Clinical Trial Data.'' \textit{Statistics in Medicine} 30: 2867--2880. 
\item
Hastie, T., Tibshirani, R. and J. Friedman (2009) \textit{The Elements of Statistical Learning}, second edition. New York: Springer.
\item
Hinkelmann, K and Kempthorme, O. (2008). \textit{Design and Analysis of Experiments} Volume I. New York: Wiley. 
\item
Holland, P. (1986) ``Statistics And Causal Inference (with discussion). \textit{Journal of the American Statistical Association} 81(396): 945--970.
\item
Imbens, G.W. (2009) ``Better LATE Than Nothing: Some Comments on Deaton (2009) and Heckman and Urzua (2009)'' \textit{Journal of Economic Literature} 48(2): 399--423.
\item
Hothorn, T., Hornik, K., and A. Zeileis (2006) ``Unbiased Recursive Partitioning: A Conditional Inference Framework.'' \textit{Journal of Computational and Graphical Statistics} 15(3): 651--674. 
\item
Johnson, V.E., Payne, R.D., Wang, T., and Mandal, S. (2016) ``On The Reproducibility of Psychological Science.'' \textit{Journal of the American Statistical Association} 112(517: 1--10. 
\item
Kass, G.V. (1980) ``An Exploratory Technique for Investigating Large Quantities of Categorical Data.'' \textit{Applied Statistics} 29(22): 119--127.
\item
Kempthorme, O. (1955) ``The Randomization Theory of Experimental Inference'' \textit{Journal of the American Statistical Association} 50: 946--967.
\item
Lee, J.D., Sun, D.L., Sun, Y., and Taylor, J.E. (2016) ``Exact Post-Selection Inference, With Application To the Lasso.'' \textit{The Annals of Statistics} 3: 907--927.
\item
Leeb, H., B.M. P\"{o}tscher (2005) ``Model Selection and Inference: Facts and Fiction,'' \textit{Econometric Theory} 21: 21--59.
\item
Leeb, H., B.M. P\"{o}tscher (2006) ``Can one Estimate the Conditional Distribution of Post-Model-Selection Estimators?''\textit{The Annals of Statistics} 34(5): 2554--2591.
\item
Leeb, H., B.M. P\"{o}tscher (2008) ``Model Selection,'' in T.G. Anderson, R.A. Davis, J.-P. Kreib, and T. Mikosch (eds.), \textit{The Handbook of Financial Time Series}, New York, Springer: 785--821.
\item
Lipkovich, I., Dmetrienko, A., Denne, J., and Enas, G. (2011) ``Subgroup Identification Based On Differential Effect Searce -- A recursive Partitioning Method For Establishing Response to Treatment In Patient Subpopulations.'' \textit{Statistics In Medicine} 30: 2601--2621.
\item
Loh, W-Y. (2011) ``Classification and Regression Trees'' \textit{WIRE's Data Mining and Discovery}, New York: Springer.
\item
Loh, W-Y, He, X., Man, M. (2015) ``A Regression Tree Approach To Identifying Subgroups With Differential Treatment Effects.'' \textit{Statistics In  Medicine} 34: 1818--1833.
\item
Meinshasen, N., Meier, L, and B\"{u}hlman (2009) ``p-Values for High-Dimensional Regression.'' \textit{Journal of the American Statistical Association} 104(48: 1671-1681.
\item
McNutt, M. (2014) ``Reproducibility.'' \textit{Science} 343 (6168): 229.
\item
Quinlan, J. R. (1986) ``Induction of Decision Trees.'' \textit{Machine Learning} 1(1): 81--106
\item
Rubin, D.B. (1986) ``Which Ifs Have Causal Answers?'' \textit{Journal of the American Statistical Association} 81: 961--962.
\item
Su, X., Tsai, C-L., Wang, H., Nickerson, D.M. and Li, B. (2009) ``Subgroup Analysis Via Recursive Partitioning.'' \textit{Journal of Machine Learning Research} 10:141--158.
\item
Swanson, S.A., Hern\'{a}n, M., Miller, M., Robins, J.M., and Richardson. (2018) ``Partial Identification of the Average Treatment Effect Using Instrumental Variables: Review of Methods for Binary Instruments, Treatments, and Outcomes.'' \textit{Journal of the American Statistical Association} DOI: 10.1080/01621459.2018.1434530.
\item
Taylor, J.E., and Tibshirani, R.J.  (2015) ``Statistical Learning And Selective Inference.'' \textit{Proceedings of the National Academy of Sciences} 112(25): 7629--7634.
\item
Thomas, M. and Bornkamp, B. (2016) ``Comparing Approaches To Treatment Effect Estimation For Subgroups In Clinical Trials.'' arXiv:1603: 03316v2 [stat.CO].
\item
Vivalt, E., (2105) ``Heterogeneous Treatment Effects in Impact Evaluation.'' \textit{American Economic Review: Papers \& Proceedings} 105(5): 467--470.
\item
Wager, S., and Athey, S. (2017) ``Estimation and Inference of Heterogeneous Treatment Effects using Random Forests.'' arXiv: 1510.04342v4 [stat.ME].
\end{description}

\end{document}